\documentclass[conference]{IEEEtran}
\IEEEoverridecommandlockouts

\usepackage[utf8]{inputenc}
\usepackage{setspace}
\usepackage[T1]{fontenc}
\usepackage[american]{babel}
\usepackage{graphicx}
\usepackage{cite}
\usepackage{listings}
\usepackage{subfig}
\usepackage{booktabs}
\usepackage{float}
\usepackage{amsmath,amssymb,exscale}

\usepackage{blindtext, graphicx}
\usepackage{verbatim}
\usepackage{multirow} 
\usepackage{algorithm}
\usepackage{algpseudocode}
\usepackage{algpseudocode}
\usepackage{fancyvrb}
\usepackage{bera}
\usepackage{amsmath}
\usepackage{mathtools}
\usepackage{lipsum}

\usepackage{times}
\usepackage{scalerel}
\usepackage{tikz}
\usetikzlibrary{svg.path}

\begin{document}
\author{
\IEEEauthorblockN{
Ruolin Du\IEEEauthorrefmark{1},
Zhiqiang Wei\IEEEauthorrefmark{1},
Zai Yang\IEEEauthorrefmark{1},
Ya-Feng Liu\IEEEauthorrefmark{2},
Bingpeng Zhou\IEEEauthorrefmark{3},
and Derrick Wing Kwan Ng\IEEEauthorrefmark{4}
}

\IEEEauthorblockA{\IEEEauthorrefmark{1}
School of Mathematics and Statistics, Xi'an Jiaotong University, Xi'an, China}

\IEEEauthorblockA{\IEEEauthorrefmark{2}
School of Mathematical Sciences, Beijing University of Posts and Telecommunications, Beijing, China}

\IEEEauthorblockA{\IEEEauthorrefmark{3}
School of Electronics and Communication Engineering, Sun Yat-sen University, Shenzhen, China}

\IEEEauthorblockA{\IEEEauthorrefmark{4}
School of Electrical Engineering and Telecommunications, University of New South Wales, Sydney, Australia}

\IEEEauthorblockA{
Email: duruolin@stu.xjtu.edu.cn,
\{zhiqiang.wei, yangzai\}@xjtu.edu.cn,
yafengliu@bupt.edu.cn,\\
zhoubp3@mail.sysu.edu.cn,
w.k.ng@unsw.edu.au
}\thanks{The work of Bingpeng Zhou was supported by the NSFC under Grant 62371478 and by Science and Technology Plan of Shenzhen under Grant JCYJ20240813151253068.}}

\title{OFDM Waveform Optimization for Bistatic Integrated Sensing and Communications}
\IEEEoverridecommandlockouts 
\maketitle
\begin{abstract}
This paper investigates the design of orthogonal frequency-division multiplexing (OFDM) waveforms for bistatic integrated sensing and communication (ISAC) systems.
In the considered framework, an ISAC transmitter jointly optimizes subcarrier assignment and power allocation for a single OFDM waveform that simultaneously supports communication and sensing functionalities. 
Meanwhile, an ISAC receiver decodes information on communication subcarriers and estimates per-path propagation delays via exploiting pilot symbols on sensing subcarriers.
We propose a joint path coefficient and delay estimation (JPCDE) scheme, revealing that the achievable communication data rate (CDR) is determined by the number of communication subcarriers, whereas the delay sensing accuracy is governed by the index distribution of sensing subcarriers.
Building on this insight, we formulate an OFDM waveform optimization problem to maximize the CDR subject to sensing-accuracy and power-budget constraints.
To solve this problem, we employ a quadratic transform and Lagrangian dual decomposition, which iteratively updates the subcarrier assignment and power allocation variables in closed-form.
Our results reveal that a subcarrier is allocated for sensing if and only if its Fisher information gain exceeds the corresponding communication rate loss, while the power allocation for communication subcarriers exhibits a bounded water-filling structure.
Simulation results demonstrate that the proposed frameworks substantially outperform existing baselines in both delay estimation accuracy and CDR.
\end{abstract}

\vspace{-0.5cm}

\section{Introduction}
The International Telecommunication Union (ITU) has identified integrated sensing and communications (ISAC) as a key enabling technology for the upcoming sixth-generation (6G) mobile communication systems \cite{ITU}. 
A fundamental challenge in ISAC lies in developing a unified transmit waveform that simultaneously supports communication and sensing while striking an effective balance between their inherently conflicting performance requirement \cite{liu2022integrated,zhang2021overview,Dong2025CommSensing,wu2025low}.
Among various candidate waveforms, orthogonal frequency-division multiplexing (OFDM) employs a set of orthonormal frequency-domain basis functions to transmit information, efficiently transforming a frequency-selective fading channel into multiple frequency-flat fading subchannels for parallel data transmission \cite{Chang1966Bell}. 
Moreover, by allocating orthogonal subcarrier groups to sensing and communication, OFDM can mitigate mutual interference between these two functionalities \cite{li2023joint}, and has been proven near-optimal for delay estimation in terms of the integral sidelobe level of the ambiguity function \cite{Liu2024OFDM}. 
Consequently, bistatic and multistatic ISAC architectures that leverage OFDM have attracted increasing attention due to their flexibility and robustness in practical deployments \cite{jiao2023information}.

A key challenge in bistatic OFDM-ISAC waveform design arises from the fundamentally different channel conditions and resource allocation on sensing and communication performance \cite{liu2022integrated,zhang2021overview}.
Specifically, for sensing, the received echo corresponding to each propagation path experiences single-path transmission, i.e., a frequency-flat fading channel. 
In contrast, the communication signal experiences multipath propagation, i.e., a frequency-selective fading channel. 
Thus, the sensing accuracy mainly depends on the index distribution of sensing subcarriers governing effective bandwidth, whereas the communication data rate (CDR) primarily depends on the quantity of communication subcarriers. 
Although existing bistatic OFDM-ISAC studies have emphasized signal processing techniques and performance analysis of sensing and communications \cite{jiao2023information,Brunner2025TMTT}, the joint waveform design problem for bistatic OFDM-ISAC remains largely underexplored.

To address these limitations, we propose a joint path coefficient and delay estimation (JPCDE) scheme and derive its Cramér–Rao bound (CRB) for delay estimation, revealing that the sensing accuracy depends on both the index distribution and power allocation of sensing subcarriers, whereas the CDR is determined by the number of communication subcarriers under frequency-selective fading.
Capitalizing on this finding, we formulate a CDR maximization problem subject to sensing accuracy and transmit power constraints, and derive closed-form iterative updates for subcarrier assignment and power allocation.
We reveal that the subcarrier assignment solution follows a principle of comparing the Fisher information gain for sensing and the communication rate loss for communications, while the power allocation for communication subcarriers exhibits a bounded water-filling structure.
Numerical results demonstrate that the proposed scheme consistently outperforms the baseline methods.

Unless otherwise specified, matrices are denoted by uppercase boldface letters and vectors are represented by lowercase boldface letters; 
$(\cdot)^T$, $(\cdot)^H$, and $(\cdot)^*$ stand for transpose, Hermitian transpose, and the complex conjugate of a matrix, respectively; 
$\boldsymbol{I}_N$ denotes the $N$-dimensional identity matrix;
$\mathbb{C}$ denotes a complex space; 
$\mathbb{E}$ denotes the expectation operator; 
$\Re(\cdot)$ and $\Im(\cdot)$ stand for the real and imaginary parts of a complex number, respectively;
$\operatorname{diag}(\cdot)$ denotes the diagonal operator;
$\boldsymbol{x}\sim \mathcal{CN}(\boldsymbol{\mu},\boldsymbol{\Sigma})$ denotes a circularly symmetric complex Gaussian vector with mean $\boldsymbol{\mu}$ and covariance matrix $\boldsymbol{\Sigma}$;
The $l_1$ norm, $l_2$ norm, and Frobenius norm are denoted by $\|\cdot\|_1$, $\|\cdot\|_2$, and $\|\cdot\|_F$, respectively.

\vspace{-0.1 cm}
\section{Bistatic OFDM-Based ISAC System Model}
\vspace{-0.1 cm}
\label{System model}
\begin{figure}[!t]
\setlength{\belowcaptionskip}{-0.5cm}
\centering
\includegraphics[width=3.3 in]{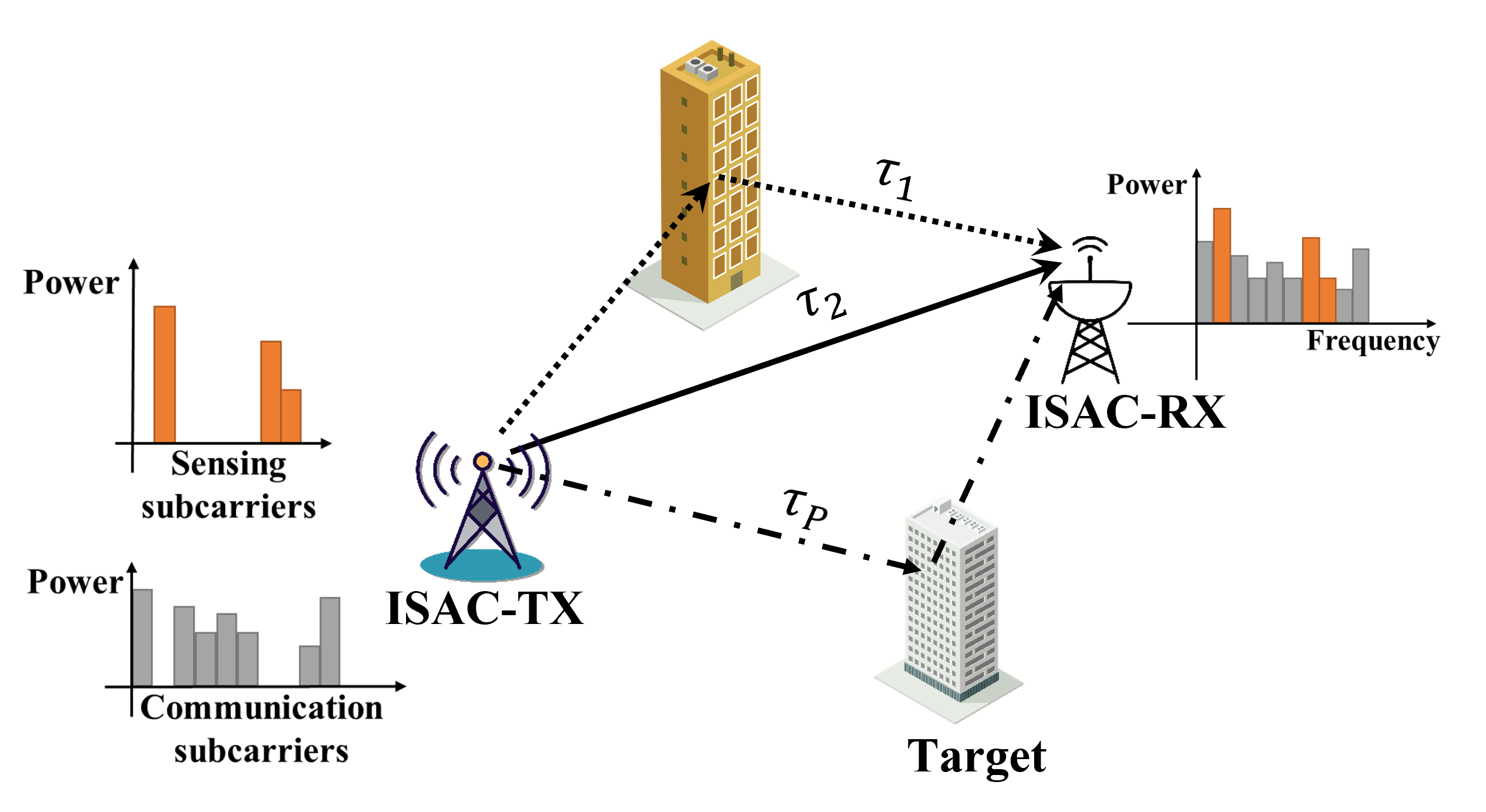}
\caption{Bistatic OFDM-based ISAC system model.}
\label{figsystemmodel}
\end{figure}
We consider a point-to-point (P2P) bistatic OFDM-based ISAC system, as depicted in Fig. \ref{figsystemmodel}. 
An ISAC transmitter (TX) emits an OFDM waveform into the environment.
After propagating through a multipath channel, an ISAC receiver (RX) aims to decode the transmitted information and estimate the delay of each path based on the received signal\footnote{Unlike a monostatic ISAC configuration, synchronization (including both time and frequency alignment) is essential due to the use of different local oscillators (LOs) and reference clocks at the TX and RX.
When both the transmitter and receiver are static and their positions are known, synchronization offsets can be well compensated leveraging existing methods, e.g., \cite{Brunner2025TMTT}.}.
Assuming that the ISAC RX is equipped with an $N_r$-element uniform linear array (ULA) with half-wavelength spacing for resolving multipath, whereas the ISAC TX adopts a single-antenna setup. 

As commonly adopted in the literature \cite{wong1999multiuser}, we assume that the CSI between the ISAC transceiver pair is known at the ISAC TX for OFDM waveform optimization, but is unknown to the ISAC RX.
Moreover, we assume that the sensing signal processing at the ISAC RX follows the pilot-based sensing scheme.
Let $f_c$, $M$, and $B$ denote the carrier frequency, the number of subcarriers, and the system bandwidth, respectively, with subcarrier spacing $\Delta f = B/ M $.
Let $\boldsymbol{u} = [u_1,\dots,u_M]^T \in \{0,1\}^M$ denote the binary subcarrier assignment vector, where $u_m$ is defined by
\begin{equation}
\small
\setlength{\abovedisplayskip}{0.1pt}
\setlength{\belowdisplayskip}{0.1pt}
\label{eq:u}
u_{m}=
    \begin{cases}
        1,& \text{if subcarrier $m$ is assigned for sensing,}\\
        0,& \text{if subcarrier $m$ is assigned for communication.}
    \end{cases}
\end{equation}
Given $\boldsymbol{u}$, $M_{\text{r}}=||\boldsymbol{u}||_1$ denotes the number of subcarriers assigned for pilot transmission and sensing, while $M_{\text{c}}=M-||\boldsymbol{u}||_1$ denotes the number of subcarriers assigned for information transmission.
Define $\boldsymbol{P} = \operatorname{diag}(P_1,\dots,P_M)$, where $P_m$, $\forall m \in \{1,\ldots,M\}$, denotes the power allocated to the $m$-th subcarrier.
The frequency-domain transmitted signal on the $m$-th subcarrier from the ISAC TX is given by ${x}_m=\sqrt{P_m}\left( u_m s_{\text{r},m} + (1 - u_m) s_{\text{c},m} \right)$, where $s_{\text{c},m}\in \mathcal{CN}(0,1)$ denotes the unknown communication data symbols.
The pilot symbol $s_{\text{r},m}$ is known to both TX and RX and assumed to be unit-modulus, i.e., $|s_{\text{r},m}|=1$ for all $m$ \cite{liu2022isaclimits,wang2024fundamentaltradeoffstimefrequencyresource}.
Assuming a finite number of static scatterers, the baseband multipath channel $\boldsymbol{h}_m\in \mathbb{C}^{N_r\times 1}$ at subcarrier $m$ is given by $\boldsymbol{h}_m=\sum_{p=1}^{P}b_p e^{-j2\pi m \Delta f \tau_p}\boldsymbol{a}_{N_r}(\psi_p^{\text{RX}})$, where $p\in \{1,\dots,P\}$ is the path index, $\tau_p \in \mathbb{R}$ denotes the propagation delay of the $p$-th path, and $b_p\in \mathbb{C}$ denotes the complex path coefficient of the $p$-th path that encapsulates both the path gain and the unknown phase noise \cite{PNkeskin2023monostaticO}.
Since the system bandwidth is much smaller than the carrier frequency, we assume that $b_p$ is constant across different subcarriers, as commonly adopted in \cite{wang2024fundamentaltradeoffstimefrequencyresource,li2023joint}.
Also, $\psi_p^{\text{RX}}$ denotes the angle-of-arrival (AoA) associated with the $p$-th path at the ISAC RX and the array steering vector is given by $\boldsymbol{a}_{N_r}(\psi)=\left[1,e^{-j\pi \cos(\psi)},\ldots,e^{-j\pi (N_r-1)\cos(\psi)}  \right]^T\in \mathbb{C}^{N_r \times 1}$.
Accordingly, the received frequency-domain signal $\boldsymbol{y}_m\in \mathbb{C}^{N_r\times 1}$ on the $m$-th subcarrier is given by $\boldsymbol{y}_m= \sqrt{P_m} \boldsymbol{h}_m \left[ u_m s_{\text{r},m} + (1 - u_m) s_{\text{c},m} \right]+ \boldsymbol{w}_m$, where $\boldsymbol{w}_m$ represents the complex additive white Gaussian noise (AWGN) vector with zero mean and covariance matrix $\sigma^2\boldsymbol{I}_{N_r}$, i.e., $\boldsymbol{w}_m\sim \mathcal{CN}(\boldsymbol{0},\sigma^2\boldsymbol{I}_{N_r})$.

When subcarrier $m$ is assigned for communication, the CDR\footnote{
As in practical OFDM systems, the CSI on the pilot-bearing (sensing) subcarriers is estimated from the received pilots and then extrapolated/interpolated to the data-bearing (communication) subcarriers for coherent detection.
This paper assumes perfect CSI extrapolation at the ISAC RX, derived from pilot-based channel estimates on the sensing subcarriers, where $R_{\text{c},m}$ serves as an upper bound on the CDR of the $m$-th communication subcarrier.} of the $m$-th subcarrier is given by $R_{\text{c},m}=(1 - u_m) \log_2 \det(\boldsymbol{I}_{N_r} + \frac{P_m}{\sigma^2}\boldsymbol{h}_m\boldsymbol{h}_m^H )$, and the overall system CDR is given by $R_\text{c} = \sum_{m=1}^{M}\left(1 - u_m\right)  \log_2 \left(1 + \frac{||\boldsymbol{h}_m||_2^2 P_m}{\sigma^2} \right)$.
It can be observed that all multipath components contribute to $||\boldsymbol{h}_m||_2^2$, indicating that the communication subsystem experiences a frequency-selective fading channel.

On the other hand, when subcarrier $m$ is assigned for sensing, the frequency-domain received sensing signal on the $m$-th subcarrier is given by $\boldsymbol{y}_{\text{r},m}=\sqrt{P_m}\sum_{p=1}^{P}b_p e^{-j2\pi m \Delta f \tau_p}\boldsymbol{a}_{N_r}(\psi_p^{\text{RX}}) u_m s_{\text{r},m}+ \boldsymbol{w}_m$.
To estimate the per-path propagation delay, the ISAC RX employs receive beamforming to separate multiple propagation paths.
To estimate the AoA of the $p$-th path, we employ spatial matched filtering, i.e., $\hat{\psi}_p^{\text{RX}}=\arg\max_{\psi}|\sum_{m=1}^M\frac{1}{\sqrt{N_r}}\boldsymbol{a}_{N_r}^H(\psi)\boldsymbol{y}_{\text{r},m}|$, as commonly adopted in the literature \cite{du2025channel}.
Assuming that $N_r$ is large and the AoAs are sufficiently separated, the received sensing signal component $y_{\text{r},m}^p$ associated with the $p$-th path at the $m$-th subcarrier can be effectively extracted based on $\hat{\psi}_p^{\text{RX}}$ and is given by $y_{\text{r},m}^p=\sqrt{P_m N_r}b_p e^{-j2\pi m \Delta f \tau_p} u_m s_{\text{r},m}+ \hat{w}_{m}^{p}$, where $\hat{w}_{m}^{p}=\frac{1}{\sqrt{N_r}}\boldsymbol{a}_{N_r}^H(\psi_p^{\text{TX}})\boldsymbol{w}_m$ denotes the residual noise after spatial matched filtering with variance $\sigma^2$. 
Furthermore, by multiplying $y_{\text{r},m}^p$ with $s_{\text{r},m}^*$, the pilot-demodulated received signal of the $p$-th path at the $m$-th subcarrier is given by
\begin{equation}
\label{eq:measurement_model}
\setlength{\abovedisplayskip}{0.1pt}
\setlength{\belowdisplayskip}{0.1pt}
\tilde{y}_{\text{r},m}^{p} = y_{\text{r},m}^{p}s_{\text{r},m}^* = \sqrt{P_m N_r} b_p  e^{-j 2\pi m \Delta f \tau_p} u_m + \tilde{w}_{m}^{p},
\end{equation}
where $\tilde{w}_{m}^{p} = \hat{w}_{m}^{p} s_{\text{r},m}^*$ remains a zero-mean circularly symmetric complex Gaussian random variable with variance $\sigma^2$, since the pilot symbols are unit-modulus.

The JPCDE method employs the maximum likelihood estimation (MLE) to jointly estimate complex path coefficients $b_p$ and delays $\tau_p$ directly from the received sensing signal in \eqref{eq:measurement_model}.
Specifically, collecting $\tilde{y}_{\text{r},m}^{p}$ in a vector $\tilde{\boldsymbol{y}}_{\text{r}}^p=[\tilde{y}_{\text{r},1}^{p},\dots,\tilde{y}_{\text{r},M}^{p} ]^T\in \mathbb{C}^{M\times 1}$ and defining the unknown parameters as $\boldsymbol{\theta}_p \triangleq [ \tau_p, \Re\{b_p\}, \Im\{b_p\}]^T$, the likelihood function to estimate $\boldsymbol{\theta}_p$ is given by
\begin{equation}
    \small
    \label{eq:likelihood}
    \setlength{\abovedisplayskip}{0.1pt}
\setlength{\belowdisplayskip}{0.1pt}
    f( \boldsymbol{\tilde{y}}_{\text{r}}^p\mid \boldsymbol{\theta}_p) =\prod_{m=1}^M \frac{1}{(\pi \sigma^2)^{\frac{u_m}{2}}} \exp\left(-\frac{| \tilde{y}_{\text{r},m}^{p} - \mu_m(\boldsymbol{\theta}_p) |^2 }{\sigma^2}\right),
\end{equation}
where $\mu_m(\boldsymbol{\theta}_p) = u_m\sqrt{P_m N_r} b_p e^{-j 2\pi m \Delta f \tau_p}$.
Note that there are $M_{\text{r}}$ nonzero entries in $\boldsymbol{\tilde{y}}_{\text{r}}^p$ and thus there are $M_{\text{r}}$ terms in the cumulative product in \eqref{eq:likelihood}.
The MLE of $\boldsymbol{\theta}_p$ is obtained as $\hat{\boldsymbol{\theta}}_p = \arg \max_{\boldsymbol{\theta}_p} f( \boldsymbol{\tilde{y}}_{\text{r}}^p\mid \boldsymbol{\theta}_p)$.

\vspace{-0.1cm}
\section{OFDM Waveform Optimization based on JPCDE}
\label{method1}
\vspace{-0.1cm}
\subsection{CRB Derivation}
\begin{figure}[!t]
\setlength{\belowcaptionskip}{-0.5cm}
\centering
\includegraphics[width=3.3 in]{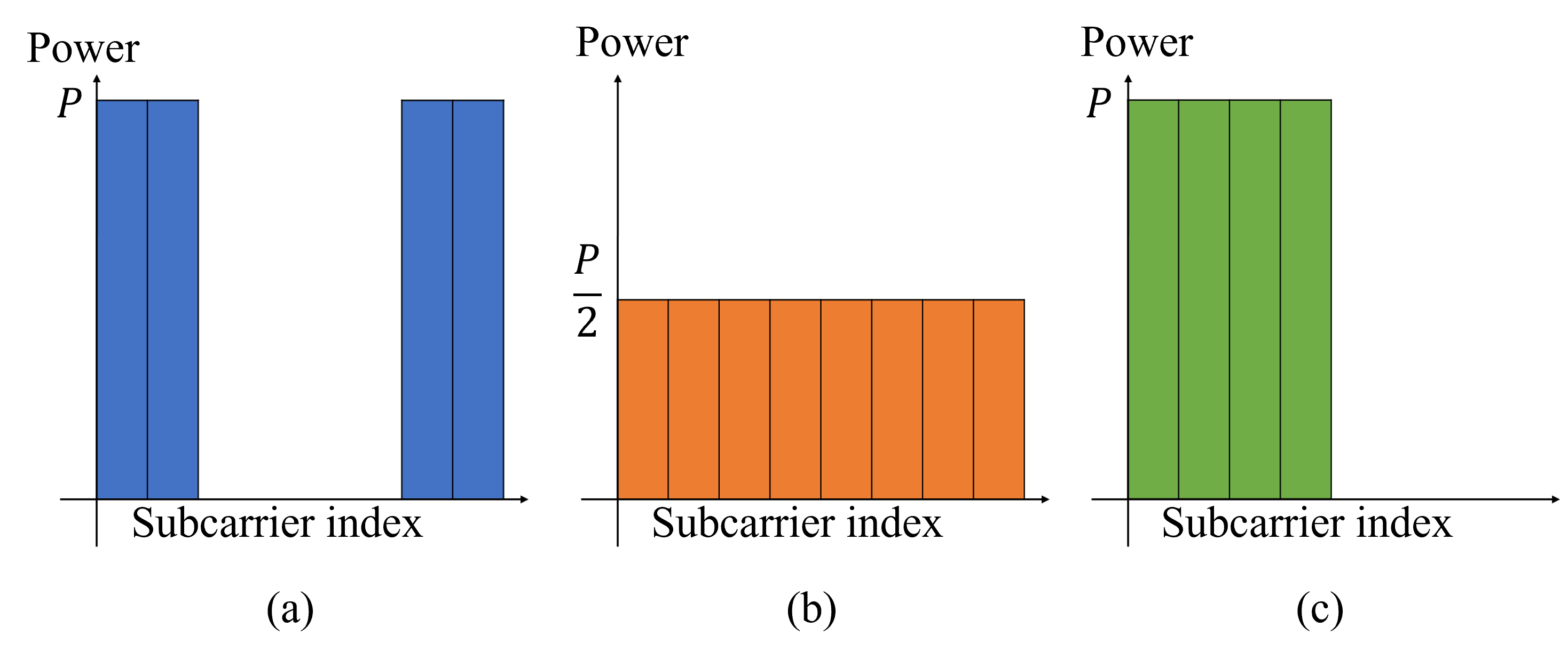}
\caption{Different sensing subcarrier assignment and power allocation strategies under an identical total power budget.}
\label{fig:EffectiveBandwidth}
\end{figure}
In this subsection, we derive the CRB for delay estimation in the proposed JPCDE scheme, which provides a fundamental lower bound on the mean-square error (MSE) of any unbiased parameter estimator \cite{kay1993estimation,wang2024fundamentaltradeoffstimefrequencyresource}.
The log-likelihood corresponding to \eqref{eq:likelihood} is given by
\begin{equation}
\small
\setlength{\abovedisplayskip}{0.1pt}
\setlength{\belowdisplayskip}{0.1pt}
\label{eq:loglikelihood}
\begin{aligned}
    \ln \left( f\left( \tilde{\boldsymbol{y}}_{\text{r}}^p \mid \boldsymbol{\theta}_p \right) \right) 
=& -\frac{\sum_{m=1}^{M} u_m}{2} \ln(\pi  \sigma^2) \\
&- \frac{1}{\sigma^2} \sum_{m=1}^{M} u_m \left| \tilde{y}_{\text{r},m}^{p} - \mu_m\left(\boldsymbol{\theta}_p\right) \right|^2.
\end{aligned}
\end{equation}
Based on \eqref{eq:loglikelihood}, the $(i,j)$-th element of the $3 \times 3$ Fisher information matrix (FIM) is given by $\boldsymbol{J}\left(\boldsymbol{\theta}_p\right)_{i,j} = - \mathbb{E} \left[ 
\frac{\partial^2 \ln \left( f\left(\boldsymbol{\tilde{y}}_{\text{r}}^p \mid \boldsymbol{\theta}_p \right) \right)}{ \partial \theta_{p,i} \, \partial \left(\theta_{p,j}\right)^* }
\right]$, where $\theta_{p,i}$ denotes the $i$-th entry of $\boldsymbol{\theta}_p$.
According to \cite{kay1993estimation}, $\boldsymbol{J}(\boldsymbol{\theta}_p)$ and the reciprocal of $\mathrm{CRB}(\tau_p)$ can be derived in closed form as \eqref{eq:FIM_explicit} and \eqref{CRB_tau}, as shown at the top of the next page.
\begin{figure*}[t]
\centering
\begin{equation}
\setlength{\abovedisplayskip}{0pt}
\setlength{\belowdisplayskip}{0pt}
\boldsymbol{J}\left(\boldsymbol{\theta}_p\right) = \frac{2N_r}{\sigma^2} \sum_{m=1}^M u_m
\begin{bmatrix}
4 \pi^2 m^2 \Delta f^2 P_m |b_p|^2 & 2\pi m \Delta f P_m \Im\{b_p\} & -2\pi m \Delta f P_m \Re\{b_p\} \\
2\pi m \Delta f P_m \Im\{b_p\} & P_m & 0 \\
-2\pi m \Delta f P_m \Re\{b_p\} & 0 & P_m
\end{bmatrix}
\label{eq:FIM_explicit}
\end{equation}
\hrulefill
\end{figure*}
\begin{figure*}[t]
\centering
\begin{equation}
\setlength{\abovedisplayskip}{0pt}
\setlength{\belowdisplayskip}{0pt}
\label{CRB_tau}
    \mathrm{CRB}^{-1}(\tau_p)= \frac{8 N_r |b_p|^2 \pi^2 \Delta f^2}{\sigma^2} \underbrace{\left( \sum_{m=1}^M P_m u_m m^2 - \frac{ \left( \sum_{m=1}^M P_m u_m m \right)^2 }{ \sum_{m=1}^M P_m u_m } \right)}_{\text{Squared effective bandwidth}}
\end{equation}
\hrulefill
\end{figure*}
\eqref{CRB_tau} reveals that the $\mathrm{CRB}(\tau_p)$ for delay estimation is inversely proportional to the squared effective bandwidth of the sensing waveform. 
The squared effective bandwidth \cite{liu2022isaclimits} is defined as the second central moment of the energy-normalized frequency distribution around its spectral centroid, quantifies the dispersion of sensing energy in the frequency domain and is jointly influenced by ${P_m}$ and ${u_m}$.
Intuitively, allocating sensing power across more widely separated subcarrier indices increases the effective bandwidth, thereby sharpening temporal resolution and improving delay estimation accuracy.
Fig.~\ref{fig:EffectiveBandwidth} compares three representative subcarrier assignment patterns under an identical total power-budget with squared effective bandwidth values of $37P$, $21P$, and $5P$, respectively.
Comparing (a) and (c), with the same number of sensing subcarriers, selecting widely separated subcarriers for sensing significantly enlarges the squared effective bandwidth and thus results in a higher delay estimation accuracy.
Moreover, comparing (a) and (b), with an identical total sensing power, scheduling fewer but more widely separated subcarriers achieves better delay estimation performance while saving more subcarriers for communication, implying a better sensing–communication tradeoff.

\subsection{Problem Formulation}
In this subsection, we formulate the OFDM waveform optimization problem that aims to maximize the system CDR, subject to individual sensing performance constraints for each path and the transmit power-budget.
The optimization problem can be formulated as:
\begin{equation}
\setlength{\abovedisplayskip}{0.1pt}
\setlength{\belowdisplayskip}{0.1pt}
\small
\begin{aligned}
\label{problem0}
& \text{OP: } \max_{\substack{\boldsymbol{P},\boldsymbol{u}
 \\}} \sum_{m=1}^{M}(1-u_{m})\log_2 \left ( 1+\frac{||\boldsymbol{h}_m||_2^2 P_m  }{\sigma^2}  \right ) \\
& \text{s.t.} \left\{ \begin{array}{l}
\text{C}1 \text{: }  \sum_{m=1}^{M} P_m \le P_{\text{req}},  \\
\text{C}2 \text{: } \sum\limits_{m=1}^{M} P_m u_m m^2- \frac{\left( \sum_{m=1}^{M} P_m u_m m\right)^2}{\sum_{m=1}^{M} P_m u_m}  
 \ge \max\limits_{p}\{ J_{\text{req},p}\},\forall p, \\
\text{C}3 \text{: }  0 \le P_m \le P_0, \forall m,
 \\
\text{C}4 \text{: }  u_m \in\left \{ 0,1 \right \} ,\forall m,
\end{array}\right.
\end{aligned}
\end{equation}
where $J_{\text{req},p}= \frac{\sigma^2 }{8 N_r |b_p|^2 \pi^2 \Delta f^2 J_{0}^2}$ and $J_{0}$ denotes the maximum tolerable per-path propagation delay estimation error.
In \eqref{problem0}, C1 represents the total power constraint, where $P_\text{req}$ denotes the transmit power-budget of ISAC TX;
C2 ensures that the per-path delay estimation CRB does not exceed the predefined sensing error threshold $J_0$;
C3 imposes a per-subcarrier power bound $P_0$ and C4 guarantees the subcarrier assignment variable to be binary.
Problem \eqref{problem0} is a mixed-integer nonlinear programming (MINLP) problem, which is generally NP-hard to solve even in the communication-only setting \cite{LiuDai2014_ComplexityOFDMA}.

First, we apply the quadratic transform \cite{shen2018fractional1} to reformulate the fractional term in C2 in \eqref{problem0} as
\begin{equation}
\setlength{\abovedisplayskip}{0.1pt}
\setlength{\belowdisplayskip}{0.1pt}
\label{FP}
\small
\frac{( \sum_{m=1}^{M} P_m u_m m)^2}{\sum_{m=1}^{M} P_m u_m}
=\max_{y} \left\{ 2y \sum_{m=1}^{M} P_m u_m m -y^2\sum_{m=1}^{M} P_m u_m\right\},
\end{equation}
where $y>0$ is an auxiliary variable.
By substituting \eqref{FP} into \eqref{problem0}, the C2 constraint is relaxed into a convex quadratic form, $\overline{\text{C}2} \text{: }\sum_{m=1}^{M} P_m u_m (m-y)^2 \ge \max_{p}\{ J_{\text{req},p}\}$.
Moreover, continuous convex relaxation is applied to replace C4 with $ \overline{\text{C}4} \text{: }0 \le u_m \le 1$ \cite{shi2021joint,boyd2004convex}.
Therefore, we transform the problem from \eqref{problem0} to the following optimization problem, which provides an upper bound for \eqref{problem0}:
\begin{equation}
\setlength{\abovedisplayskip}{0.1pt}
\setlength{\belowdisplayskip}{0.1pt}
\begin{aligned}
\label{problem1}
& \text{OP': } \max_{\substack{ \boldsymbol{P},\boldsymbol{u}
 ,y\\}} \sum_{m=1}^{M}(1-u_{m})\log_2 \left ( 1+\frac{||\boldsymbol{h}_m||_2^2 P_m  }{\sigma^2}  \right ) \\
& \text{s.t.} \left\{ \begin{array}{l}
\overline{\text{C}2} \text{: }  \sum_{m=1}^{M} P_m u_m (m-y)^2 \ge \max_{p}\{ J_{\text{req},p}\}, \\
\text{C1, C3},\\
\overline{\text{C}4} \text{: }  0 \le u_m \le 1, \forall m.
\end{array}\right.
\end{aligned}
\end{equation}
In \eqref{problem1}, following \cite{shen2018fractional1}, the auxiliary variable $y$ admits a closed-form optimal solution when both $\boldsymbol{P}$ and $\boldsymbol{u}$ are fixed
\begin{equation}
\setlength{\abovedisplayskip}{0.1pt}
\setlength{\belowdisplayskip}{0.1pt}
 y^*=\frac{\sum_{m=1}^M P_m u_m m}{\sum_{m=1}^M P_m u_m}.
\label{eq:y_update}
\end{equation}
$y^*$ in \eqref{eq:y_update} represents the power-weighted spectral centroid of the subcarriers allocated for sensing, while the constraint $\overline{\text{C}2}$ in \eqref{problem1} is the corresponding power-weighted spectral variance around this centroid.
Moreover, when $y$ is fixed, the subproblems of optimizing $\boldsymbol{u}$ for given $\boldsymbol{P}$ or optimizing $\boldsymbol{P}$ for given $\boldsymbol{u}$ are convex. 
Therefore, we propose a block coordinate descent (BCD) algorithm capitalizing the Lagrangian dual decomposition to obtain a suboptimal solution for \eqref{problem1}.

\subsection{Solution Approach}
The Lagrangian of \eqref{problem1} is given by \eqref{L} at the top of the next page.
\begin{figure*}[t]
\centering
\begin{equation}
\setlength{\abovedisplayskip}{0pt}
\small
\label{L}
\mathcal{L}(\boldsymbol{P},\boldsymbol{u},y,\lambda,\mu)=\sum_{m=1}^M(1-u_m) \log_2 \left(1+\frac{||\boldsymbol{h}_m||_2^2  P_m}{\sigma^2}\right) - \lambda\left( \sum_{m=1}^{M}P_m -P_{\mathrm{req}}\right) + \mu\left(\sum_{m=1}^M P_m u_m(m-y)^2-\max_{p}\{ J_{\text{req},p}\}\right)
\end{equation}
\hrulefill
\end{figure*}
Here, $\lambda \geq 0$ and $\mu \geq 0$ are the Lagrange multipliers associated with constraints C1 and $\overline{\text{C}2}$ in \eqref{problem1}, respectively.
Meanwhile, the boundary constraints C3 and $\overline{\text{C}4}$ are incorporated into the Karush-Kuhn-Tucker (KKT) conditions in the subsequent analysis.
The dual problem of \eqref{problem1} is given by
\begin{equation}
\setlength{\abovedisplayskip}{0.1pt}
\setlength{\belowdisplayskip}{0.1pt}
    \min_{\lambda \ge0,\mu \ge 0} \sup_{\boldsymbol{P},\boldsymbol{u},y} \mathcal{L}(\boldsymbol{P},\boldsymbol{u},y,\lambda,\mu).
\end{equation}
The overall optimization procedure involves alternating updates of $\boldsymbol{P}, \boldsymbol{u}$, and the dual variables $\lambda$ and $\mu$.
In the following, we assume that the superscript $(\cdot)^{(k)}$ is adopted to indicate the $k$-th iteration of the alternating iterative updates.

\subsubsection{Update $\boldsymbol{P}^{(k)}$ and $y^{(k)}$ with fixed $\boldsymbol{u}^{(k)}$}
Since $u_m^{(k)} \in \{0,1\}, \forall m$, \eqref{L} can be decoupled into terms associated exclusively with sensing subcarriers and communication subcarriers\footnote{Although $u_m$ is relaxed to a continuous variable in $[0,1]$, we will prove in Section~\ref{Updateu} that the optimal solution of $u_m$ always lies on the boundary of the feasible set, i.e., $u_m \in \{0,1\}$, ensuring that each subcarrier is assigned exclusively to either sensing or communication.}.
By applying Lagrangian dual decomposition,  \eqref{L} can be decomposed as $\mathcal{L}^{(k)} =  \mathcal{L}_{\text{c}}^{(k)} +  \mathcal{L}_{\text{r}}^{(k)} + \lambda^{(k)} P_{\mathrm{req}} - \mu^{(k)} \max_{p}\{ J_{\text{req},p}\}$, where the communication-related (i.e., $u_m^{(k)}=0$) Lagrangian contribution is given by
\begin{equation}
\setlength{\abovedisplayskip}{0.1pt}
\setlength{\belowdisplayskip}{0.1pt}
\small
\label{eq:Lc}
    \mathcal{L}_{\text{c}}^{(k)} \triangleq \sum_{m=1}^M (1-u_m^{(k)})\log_2 (1+\frac{||\boldsymbol{h}_m||_2^2 P_m}{\sigma^2})-\lambda^{(k)}\sum_{m=1}^M (1-u_m^{(k)})P_m,
\end{equation}
and the sensing-related (i.e., $u_m^{(k)}=1$) Lagrangian contribution is given by
\begin{equation}
\setlength{\abovedisplayskip}{0.1pt}
\setlength{\belowdisplayskip}{0.1pt}
\small
\label{eq:Lr}
    \mathcal{L}_{\text{r}}^{(k)} \triangleq \mu^{(k)} \sum_{m=1}^M  u_m^{(k)} P_m (m-y^{(k)})^2-\lambda^{(k)} \sum_{m=1}^M  u_m^{(k)}P_m.
\end{equation}

For subcarriers assigned for communication (i.e., $u_m^{(k)} = 0$), the objective function is given by \eqref{eq:Lc}.
By exploiting the KKT optimality conditions \cite{wong1999multiuser,boyd2004convex}, the optimal power allocation on the $m$-th subcarrier used for communication is given by:
\begin{equation}
\setlength{\abovedisplayskip}{0.1pt}
\setlength{\belowdisplayskip}{0.1pt}
\label{eq:WF}
  P_m^{(k)}= \left[ \frac{1}{\lambda^{(k)} \ln 2}-\frac{\sigma^2}{||\boldsymbol{h}_m||_2^2} \right]_0^{P_0},
\end{equation}
where $[x]_a^b=\min \{b,\max\{x,a\}\}$.
The power allocation solution in \eqref{eq:WF} follows a bounded water-filling structure, where more power is allocated to subcarriers with higher channel conditions, i.e., those exhibiting higher channel gains $||\boldsymbol{h}_m||_2^2$.

For subcarriers assigned for sensing (i.e., $u_m^{(k)} = 1$), the objective function is given by \eqref{eq:Lr}.
The partial derivative of $\mathcal{L}_{\text{r}}^{(k)}$ with respect to (w.r.t.) $P_m$ is given by
\begin{equation}
\setlength{\abovedisplayskip}{0.1pt}
\setlength{\belowdisplayskip}{0.1pt}
\label{eq:sensingPower}
    \frac{\partial \mathcal{L}_{\text{r}}^{(k)}}{\partial P_m} =\mu^{(k)} (m-y^{(k)})^2 -\lambda^{(k)}.
\end{equation}
The term $(m-y^{(k)})^2$ measures the squared distance in subcarrier index space between subcarrier $m$ and the power-weighted spectral centroid $y^{(k)}$.
Moreover, as $\mathcal{L}_{\text{r}}^{(k)}$ is linear w.r.t. $P_m$, a positive gradient in \eqref{eq:sensingPower} implies that the optimal power allocation is full power transmission $P_m^{(k)}=P_0$ for those with $\frac{\partial \mathcal{L}_{\text{r}}^{(k)}}{\partial P_m} >0$.
To ensure feasible power allocation under the total power budget C1 in \eqref{problem1}, we adopt a power allocation strategy for sensing subcarriers to maximize $\mathcal{L}_{\text{r}}^{(k)}$.
Specifically, we first sort the sensing subcarriers in descending order of $(m-y^{(k)})^2$ values, and re-index them as $(m_1-y^{(k)})^2 \ge \dots \ge (m_q-y^{(k)})^2 \dots \ge(m_{M_{\text{r}}}-y^{(k)})^2\ge 0 $.
Then, define $\bar{Q}^{(k)}\triangleq\min \left\{Q: \sum_{q=1}^{Q} (m_q-y^{(k)})^2 P_0 \ge \max_{p}\{ J_{\text{req},p}\}  \right\}$ as the minimal number of subcarriers (with full power $P_0$) needed to satisfy the CRB threshold in $\overline{\text{C}2}$ in \eqref{problem1}.
Then, the closed-form optimal power allocation $P_{m_q}^{(k)}$ for the sorted sensing subcarriers is given by
\begin{equation}
\setlength{\abovedisplayskip}{0.1pt}
\setlength{\belowdisplayskip}{0.1pt}
\small
P_{m_q}^{(k)} =
\begin{cases}
P_0, & \text{if } q < \bar{Q}^{(k)}, \\[6pt]
\displaystyle \frac{\max\limits_{p}\{ J_{\text{req},p}\} - \sum_{q=1}^{\bar{Q}^{(k)}-1}  (m_q-y^{(k)})^2 P_0}{ (m_{_{\bar{Q}^{(k)}}}-y^{(k)})^2}, 
& \begin{array}[t]{@{}l@{}} 
\text{if } q = \bar{Q}^{(k)}, 
\end{array} \\[8pt]
0, & \text{if } q > \bar{Q}^{(k)},
\end{cases}
\label{eq:P_update}
\end{equation}
This implies that subcarriers farther from $y^{(k)}$ tend to be allocated higher power to maximize the delay estimation accuracy.
In practical implementation, if the combined power allocation from \eqref{eq:WF} and \eqref{eq:P_update} violates the total power constraint C1, the problem becomes infeasible. 
To resolve this, the dual variable $\lambda^{(k)}$ is increased in subsequent iterations, thereby imposing a stronger penalty on excessive power usage in the Lagrangian.
When $\bar{Q}^{(k)} > M_{\text{r}}$, the sensing constraint is infeasible with the current sensing subcarrier solution.
To address this, $\mu^{(k)}$ is increased to enhance the incentive for subcarrier reassignment from communication to sensing in \eqref{eq:L_m^k}.
Given the current sensing power allocation in \eqref{eq:P_update} and the subcarrier allocation variable $\boldsymbol{u}^{(k)}$, the auxiliary variable $y$ is updated via \eqref{eq:y_update}. 

\subsubsection{Update $\boldsymbol{u}^{(k+1)}$ with fixed $\boldsymbol{P}^{(k)}$ and $y^{(k)}$}
\label{Updateu}
The partial derivative of the Lagrangian function $\mathcal{L}^{(k)}$ in \eqref{L} w.r.t. $u_m$ is given by
\begin{equation}
\setlength{\abovedisplayskip}{0.1pt}
\setlength{\belowdisplayskip}{0.1pt}
\label{eq:L_m^k}
\small
\begin{aligned}
       \mathcal{L}_m^{(k)} = \underbrace{- \log_2 \left(1+\frac{||\boldsymbol{h}_m||_2^2 P_m^{(k)}}{\sigma^2}\right)}_{\text{CDR loss}}+\underbrace{\mu^{(k)}P_m^{(k)}(m-y^{(k)})^2}_{\text{Sensing Fisher information gain}}.
\end{aligned}
\end{equation}
The first term in \eqref{eq:L_m^k} represents the CDR loss when subcarrier $m$ is assigned for sensing, while the second term represents the corresponding sensing Fisher information gain.
Critically, even when the binary constraint $u_m \in \{0,1\}$ is relaxed to $u_m \in [0,1]$, the optimal solution remains binary. 
This occurs because the derivative in \eqref{eq:L_m^k}
is independent of $u_m^{(k)}$, making the Lagrangian $\mathcal{L}^{(k)}$ itself in \eqref{L} linear w.r.t. $u_m^{(k)}$.
Accordingly, the closed-form optimal update rule for $\boldsymbol{u}^{(k+1)}$ is given by
\begin{equation}
\setlength{\abovedisplayskip}{0.1pt}
\setlength{\belowdisplayskip}{0.1pt}
\label{eq:u_update}
   u_m^{(k+1)}= \begin{cases}1, & \text { if } \mathcal{L}_m^{(k)} \ge 0, \\ 0, & \text { otherwise .}\end{cases} 
\end{equation}
The above proposed subcarrier assignment rule offers two advantages compared to existing approaches \cite{shi2021joint,li2023joint}.
First, it provides a physically interpretable decision criterion: a subcarrier is assigned to sensing if and only if its Fisher information gain exceeds the corresponding communication rate loss, which provides an intuitive characterization of the fundamental sensing–communication tradeoff.
Second, our formulation ensures the optimal solution for $u_m^{(k)}$ is binary, as required in the original formulation \eqref{problem0}.

\subsubsection{Update dual variables $\lambda^{(k+1)}$ and $\mu^{(k+1)}$}
The dual variables are updated leveraging the subgradient method:
{\small
\setlength{\abovedisplayskip}{0.1pt}%
\setlength{\belowdisplayskip}{0.1pt}%
\begin{align}
\lambda^{(k+1)}
&= [\lambda^{(k)}+\eta_{\lambda}^{(k)}(\sum_{m=1}^M P_m^{(k)} -P_{\mathrm{req}})]^{+}, \label{eq:lambda_update}\\[-4pt]
\mu^{(k+1)}
&= [\mu^{(k)}+\eta_{\mu}^{(k)}(\max\limits_{p}\{ J_{\text{req},p}\}-\sum_{m=1}^M P_m^{(k)} u_m^{(k)}(m-y^{(k)})^2)]^{+}, \label{eq:mu_update}
\end{align}}where $\eta_{\lambda}$ and $\eta_{\mu}$ denote the step sizes, and $[\cdot]^+$ denotes projection onto the nonnegative orthant.

The convergence of the proposed JPCDE-based OFDM waveform optimization algorithm is established as follows. 
First, the relaxed problem \eqref{problem1} maintains equivalence with \eqref{problem0} through the quadratic transformation of $\overline{\text{C}2}$ \cite{shen2018fractional1} and the binary preservation of $u_m$ via \eqref{eq:u_update}, ensuring solution feasibility. 
Second, the BCD procedure generates a nondecreasing sequence of Lagrangian values $\mathcal{L}^{(k)}$, while proper step sizes in \eqref{eq:lambda_update} and \eqref{eq:mu_update} guarantee convergence to a coordinate-wise stationary point of the Lagrangian \cite{boyd2004convex}.
The computational complexity of the waveform optimization in JPCDE is dominated by the sorting of sensing subcarriers in \eqref{eq:sensingPower}, yielding an overall complexity of $\mathcal{O}(M \log M)$ per iteration.

\vspace{-0.3 cm}
\section{Simulation Results}
\begin{figure}[!t]
\setlength{\belowcaptionskip}{-0.5cm}
\centering
\includegraphics[width=3.3 in]{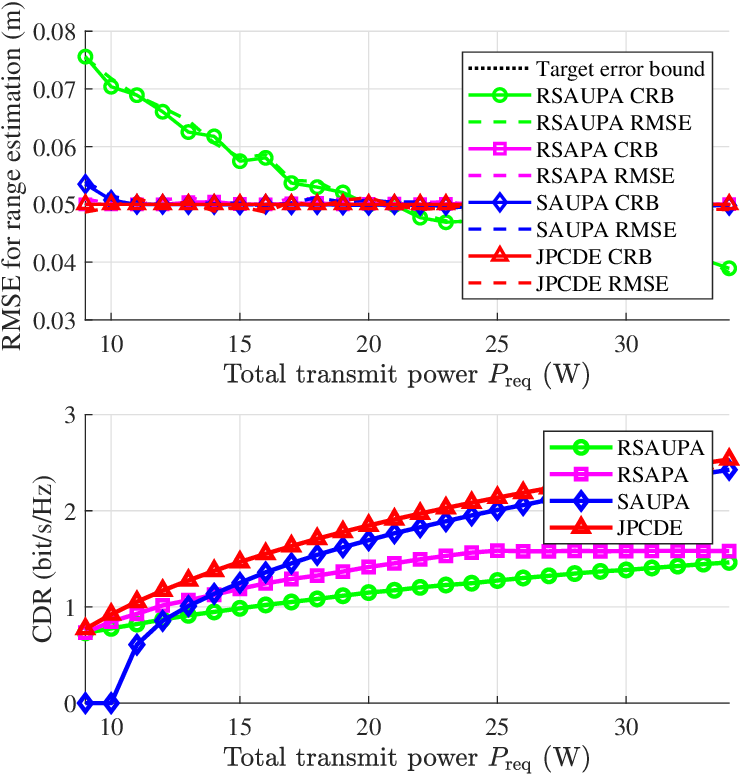}
\caption{Sensing and communication performance versus $P_{\text{req}}$ of the proposed and baseline schemes, with range error bound set to $0.05~\text{m}$.}
\label{fig:method1_comparison}
\end{figure}

\vspace{-0.2 cm}
\label{simulationresult}
We consider a system employing $M = 1024$ subcarriers, and the number of propagation paths is $P = 6$. 
We set $\Delta f = 150$ kHz, $N_r = 16$, $\sigma^2 = 1 \times 10^{-3}$ W, $P_0 = 4 \times 10^{-2}$ W, and $c = 3 \times 10^8$ m/s.
The results are averaged over 3,000 Monte Carlo simulation trials.
In addition, the delay estimation error is converted into the corresponding range error by multiplying by the speed of light 
$c$, providing a more intuitive measure of sensing accuracy.
\begin{figure}[!t]
\setlength{\belowcaptionskip}{-0.5cm}
\centering
\includegraphics[width=3.3 in]{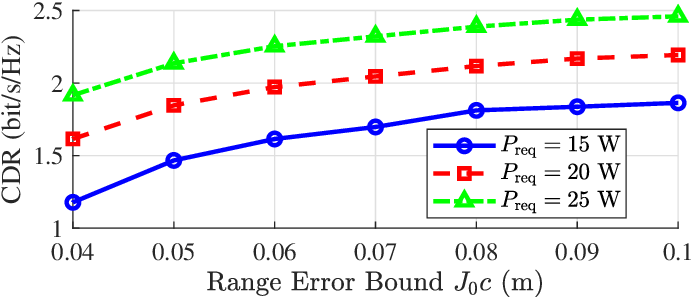}
\caption{CDR versus the range error bound under different power budgets for the proposed JPCDE-based waveform optimization.}
\label{fig:method1_sumrate_vs_rangeErr}
\end{figure}
Fig.~\ref{fig:method1_sumrate_vs_rangeErr} illustrates the tradeoff between the system CDR and the sensing accuracy for the proposed waveform optimization method. 
As the range error bound increases, the achievable CDR increases. 
This is because a larger range error bound corresponds to a more relaxed sensing constraint, thereby permitting more power and subcarriers to be allocated for communication.
Moreover, a higher transmit power-budget $P_{\text{req}}$ results in improved CDR across all range-error bounds, i.e., a better communication-sensing tradeoff, as the ISAC TX enjoys a higher flexibility in power allocation.
Fig.~\ref{fig:method1_comparison} compares the proposed scheme against three baseline approaches, as considered in \cite{shi2021joint,li2023joint}: subcarrier assignment with uniform power allocation (\textbf{SAUPA}), random subcarrier assignment with power allocation (\textbf{RSAPA}), and random subcarrier assignment with uniform power allocation (\textbf{RSAUPA})\footnote{In both RSAPA and RSAUPA, $50\%$ of the subcarriers are randomly selected for sensing.}.
The proposed JPCDE scheme and the RSAPA scheme maintain both the CRB and empirical root mean squared error (RMSE) of range estimation consistently around the target error bound of $0.05$ m across all transmit power-budgets, indicating a feasible waveform design. 
However, JPCDE allocates fewer sensing subcarriers than RSAPA and thus JPCDE achieves a much better CDR than RSAPA.
Moreover, the SAUPA scheme satisfies the sensing accuracy requirement when $P_{\text{req}}\geq11$ W, and achieves a CDR close to JPCDE and higher than RSAPA, while the RSAUPA scheme performs the worst in terms of both communication and sensing.
This difference underscores the importance of subcarrier assignment in waveform optimization, particularly with uniform power allocation.

\vspace{-0.2 cm}
\section{Conclusion}
\label{conclusion}
This paper investigated the waveform optimization problem for a bistatic OFDM-based ISAC system.
We proposed a JPCDE scheme and derived its delay estimation CRB, revealing that the sensing accuracy relies on sensing subcarrier index distribution, while the CDR is governed by the number of communication subcarriers and the channel's frequency-selective fading. 
Simulation results demonstrated that the proposed scheme achieved centimeter-level range sensing accuracy under practical power constraints and outperformed conventional baselines in terms of both sensing and communication.
\bibliographystyle{IEEEtran}
\bibliography{main}

@misc{ITU,
  author       = {{ITU}},
  title        = {{Introduction to (6G) | IMT-2030}},
  howpublished = {\url{https://www.tonex.com/training-courses/introduction-to-6g-imt-2030/}},
  year         = {2023},
  note         = {Online; accessed Sep. 26, 2025}
}

@book{boyd2004convex,
  author    = {S. P. Boyd and L. Vandenberghe},
  title     = {Convex Optimization},
  publisher = {Cambridge Univ. Press},
  address   = {Cambridge, U.K.},
  year      = {2004}
}

@ARTICLE{LiuDai2014_ComplexityOFDMA,
  author    = {Y.-F. Liu and Y.-H. Dai},
  title     = {On the Complexity of Joint Subcarrier and Power Allocation for Multi‑User {OFDMA} Systems},
  journal   = {IEEE Trans. Signal Process.},
  volume    = {62},
  number    = {3},
  pages     = {583--596},
  month     = {Feb.},
  year      = {2014},
  doi       = {10.1109/TSP.2013.2293130}
}

@article{jiao2023information,
    author    = {T. Jiao and others},
  title     = {Information-Theoretic Limits of Bistatic Integrated Sensing and Communication},
  journal   = {IEEE Trans. Inf. Theory},
  pages     = {9302--9318},
  volume    = {71},
  number    = {12},
  year      = {2025},
  month     = {Dec.},
  doi       = {10.1109/TIT.2025.3621465}
}

@ARTICLE{wu2025low,
  author    = {J. Wu and others},
  title     = {Low-altitude wireless networks: A comprehensive survey},
  journal   = {arXiv preprint arXiv:2509.11607},
  year      = {2025},
  month     = {Sep.}
}

@ARTICLE{Liu2024OFDM,
  author    = {Z. Du and others},
  title     = {Reshaping the {ISAC} Tradeoff Under {OFDM} Signaling: A Probabilistic Constellation Shaping Approach},
  journal   = {IEEE Trans. Signal Process.},
  volume    = {72},
  pages     = {4782--4797},
  year      = {2024},
  month     = {Sep.},
  doi       = {10.1109/TSP.2024.3465499}
}

@ARTICLE{Chang1966Bell,
  author    = {R. W. Chang},
  title     = {Synthesis of Band-Limited Orthogonal Signals for Multichannel Data Transmission},
  journal   = {Bell Syst. Tech. J.},
  volume    = {45},
  number    = {10},
  pages     = {1775--1796},
  year      = {1966},
  month     = {Dec.},
  doi       = {10.1002/j.1538-7305.1966.tb02435.x}
}

@article{PNkeskin2023monostaticO,
  author    = {M. F. Keskin and others},
  title     = {Monostatic Sensing With {OFDM} Under Phase Noise: From Mitigation to Exploitation},
  journal   = {IEEE Trans. Signal Process.},
  volume    = {71},
  pages     = {1363--1378},
  year      = {2023},
  month     = {Jan.},
  doi       = {10.1109/TSP.2023.3266976}
}

@article{wang2024fundamentaltradeoffstimefrequencyresource,
      title={On the Fundamental Trade-Offs of Time-Frequency Resource Distribution in {OFDMA ISAC}}, 
      author={Xiao-Yang Wang and others},
    journal   = {arXiv preprint arXiv:2407.12628 },
    year      = {2024},
    month     = {Jul.},
}

@article{shi2021joint,
  author    = {C. Shi and others},
  title     = {Joint Optimization Scheme for Subcarrier Selection and Power Allocation in Multicarrier Dual-Function Radar-Communication System},
  journal   = {IEEE Syst. J.},
  volume    = {15},
  number    = {1},
  pages     = {947--958},
  year      = {2021},
  month     = {Apr.},
  doi       = {10.1109/JSYST.2020.2984637}
}

@article{li2023joint,
  author    = {Y. Li and others},
  title     = {Joint Subcarrier and Power Allocation for Uplink Integrated Sensing and Communication System},
  journal   = {IEEE Sens. J.},
  volume    = {23},
  number    = {24},
  pages     = {31072--31081},
  month     = {Dec.},
  year      = {2023},
  doi       = {10.1109/JSEN.2023.3330936}
}

@ARTICLE{Brunner2025TMTT,
  author    = {D. Brunner and others},
  title     = {Bistatic {OFDM}-Based {ISAC} With Over-the-Air Synchronization: System Concept and Performance Analysis},
  journal   = {IEEE Trans. Microw. Theory Techn.},
  volume    = {73},
  number    = {5},
  pages     = {3016--3029},
  year      = {2025},
  month     = {May},
  doi       = {10.1109/TMTT.2024.3487295}
}

@article{wong1999multiuser,
  author    = {C. Y. Wong and others},
  title     = {Multiuser {OFDM} with Adaptive Subcarrier, Bit, and Power Allocation},
  journal   = {IEEE J. Sel. Areas Commun.},
  volume    = {17},
  number    = {10},
  pages     = {1747--1758},
  year      = {1999},
  month     = {Oct.},
  doi       = {10.1109/49.793310}
}

@ARTICLE{liu2022isaclimits,
  author    = {A. Liu and others},
  title     = {A Survey on Fundamental Limits of Integrated Sensing and Communication},
  journal   = {IEEE Commun. Surv. Tut.},
  year      = {2022},
  volume    = {24},
  number    = {2},
  pages     = {994--1034},
  doi       = {10.1109/COMST.2022.3149272}
}

@ARTICLE{du2025channel,
  author    = {R. Du and others},
  title     = {Channel Knowledge Map-assisted Dual-domain Tracking and Predictive Beamforming for High-Mobility Wireless Networks},
  journal={IEEE Trans. Wireless Commun.}, 
  year={2026},
  volume={25},
  pages={10968--10985},
  month= {Jan.}
}

@article{liu2022integrated,
  title = {Integrated Sensing and Communications: Toward Dual-Functional Wireless Networks for {6G} and Beyond},
  author = {Liu, Fan and others},
  year = {2022},
  month = jun,
  journal = {IEEE J. Sel. Areas Commun.},
  volume = {40},
  number = {6},
  pages = {1728--1767},
  issn = {0733-8716, 1558-0008},
  doi = {10.1109/JSAC.2022.3156632},
  urldate = {2022-09-07}
}

@book{kay1993estimation,
  author    = {S. M. Kay},
  title     = {Fundamentals of Statistical Signal Processing: Estimation Theory},
  publisher = {Prentice-Hall, Inc.},
  year      = {1993},
  address   = {Englewood Cliffs, NJ, USA}
}

@article{zhang2021overview,
  author={Zhang, J. Andrew and others},
  journal={IEEE J. Sel. Top. Signal Process.}, 
  title={An Overview of Signal Processing Techniques for Joint Communication and Radar Sensing}, 
  year={2021},
  month = nov,
  volume={15},
  number={6},
  pages={1295-1315},
  doi={10.1109/JSTSP.2021.3113120}}

@article{shen2018fractional1,
  author    = {K. Shen and W. Yu},
  title     = {Fractional Programming for Communication Systems—Part {I}: Power Control and Beamforming},
  journal   = {IEEE Trans. Signal Process.},
  volume    = {66},
  number    = {10},
  pages     = {2616--2630},
  year      = {2018},
  month     = {May},
  doi       = {10.1109/TSP.2018.2812733}
}

@ARTICLE{Dong2025CommSensing,
  author    = {F. Dong and others},
  title     = {Communication-Assisted Sensing in {6G} Networks},
  journal   = {IEEE J. Sel. Areas Commun.},
  volume    = {43},
  number    = {4},
  pages     = {1371--1386},
  month     = {Apr.},
  year      = {2025},
  doi       = {10.1109/JSAC.2025.3531548},
}
\end{document}